# A Time-shared Photonic Reservoir Computer for Big Data Analytics

Dharanidhar Dang, *Member, IEEE*, Rabi Mahapatra, Senior *Member, IEEE*

*Abstract—* **Information processing has reached the era of big data. Big data challenges are difficult to address with traditional Von Neumann or Turing approach. Hence implementation of new computational techniques is highly essential. Nanophotonics with its remarkable speed and multiplexing capability is a promising candidate for such implementations. This paper proposes a novel photonic computing system made-up of Mach-Zehnder interferometer and an optical fiber spool to emulate a powerful machine learning technique called reservoir computing. The proposed system is also integrated with a time-division-multiplexing circuit to facilitate parallel computation of multiple tasks which is first of its kind. The proposed design performs large-scale tasks like spoken digit recognition, channel equalization, and time-series prediction. Experimental results with standard photonic simulator demonstrate significant performance in terms of speed and accuracy compared to state of the art digital and software implementations.**

## I. Introduction

Big data computing demands ultrafast computation with least amount of power consumption which the conventional high performance systems won't offer. Recently, brain processing dynamics has inspired research communities to propose several novel computational concepts [1,2] to solve big data problems. One such technique is reservoir computing (RC) which is also known as liquid state machine. RC is based on the computational power of complex recurrent neural networks (RNNs) operating in a dynamical and transient-like fashion. In a typical recurrent neural network, it is very complex and time consuming to train network connection weights. However, RC does not need training of its network weights; only the weights connected to the output layers are trained. Thus, the overall computing time in RC is lower than a traditional RNN. Fig.1 shows a theoretical illustration of the network structure typically considered in RC. Th details of RC is given in the next section. Due to large number of dynamical elements, big data applications such as complex classification tasks and nonlinear approximations can be effectively realized with RC.

A software-based RC implementation on a traditional computer offers limited performance. Silicon photonics based RC implementation appears to be a promising alternative to address this problem. Over the years, several photonic RC implementations have been proposed to achieve high performance computing [1-3] for big data. However, all these photonic implementations execute one task at a time. To achieve high order performance, and utilize the photonic infrastructure, multitasking features should be integrated into the photonic RC system. This motivates us to design a new photonic RC system that offers multitasking to tackle big data applications. This paper introduces a time-shared RC systems where in time-division-multiplexing (TDM) scheme is integrated with reservoir node for computation. This paper has following contributions:

- A single node photonic reservoir integrated with time-division-multiplexing (TDM) scheme is proposed. It allows multiple tasks to be computed in a time-shared fashion on a single RC. We believe this is the first time such an attempt is made for for photonic RC.
- The proposed time-shared photonic RC design is simulated using appropriate tools and benchmarks were run to demonstrate efficacy of the proposed approach. Simulation results show prediction error = $0.035 \pm 0.001$ for Santa Fe time series prediction, prediction error= $1.38 \times 10^{-4.5}$ for non-linear channel equalization, prediction error=0.08% for spoken digit recognition, and prediction error=0.08% for NARMA task, when all these tasks were computed in a time-shared manner.

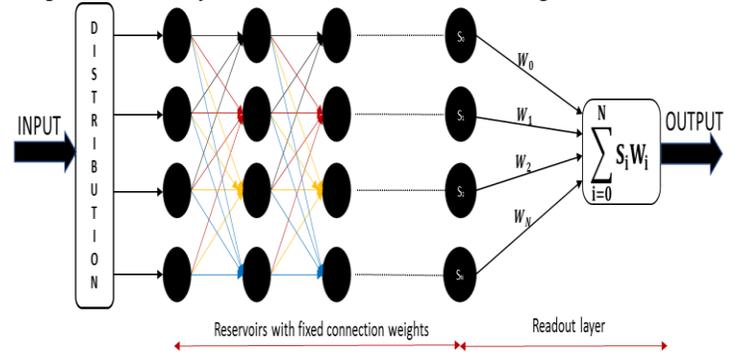

Fig.1: Conceptual representation of Reservoir Computing

The paper is organized as follows. Details of the proposed photonic reservoir architecture is depicted in Section II. Section III presents the design space exploration of the proposed system. Experiments, results, and comparative analysis are presented in Section IV followed by conclusion.

## II. Photonic Reservoir Computing

RC is increasingly being in use in the field of data science due to its high-speed prediction mechanism. The working principle of RC is as follows.

### A. Principle of Reservoir Computing

A RC comprises of an input layer, a reservoir layer, and an output layer as shown in Fig.1. The input layer of a RC distributes input data to its reservoir layer in discrete time

Dharanidhar Dang is with Texas A&M university, College Station, TX USA (e-mail: d.dharanidhar@tamu.edu).

Rabi Mahapatra is with Texas A&M university, College Station, TX (e-mail: rabi@tamu.edu).

through fixed connection weights. The reservoir layer is a dynamical system whose state at discrete time step *n* can be described as a set of *N* scalar variables $S_i(n)$ $(i = 1,2,...N)$ called neurons. All the neurons in a reservoir layer are randomly interconnected with fixed weights, constituting a recurrent network (i.e. a network of neurons having feedback loops). Under the influence of input data, the reservoir layer exhibits transient responses. The transient behavior of a reservoir is governed by an evolution equation as depicted in Equation.1.

$$S_i(n) = f[\propto C_i x(n) + \beta \sum_1^N w_{ij} s_j(n-1)] \quad (1)$$

Here $S_i(n)$ is the state of $i^{th}$ neuron at discrete time $n$, $f$ is a non-linear function, $x(n)$ is input to RC at discrete time $n$, $C_i$ & $w_{ij}$ are the connection coefficients that define the topology of a reservoir layer, and $\propto$ & $\beta$ are tuning parameters to regulate the dynamics of a reservoir. The transient states $S_i(n)$ are fed to the output layer through readout weights $W_i$ to determine the output $O(n)$.

$$O(n) = \sum_1^N W_i s_i(n) + W_{bias} \quad (2)$$

Here $W_{bias}$ is the bias value required for the training of RC. During the training phase, the readout-layer weights $W_i$ and bias weight $W_{bias}$ are optimized to minimize error between the expected output $O'(n)$ and the actual output $O(n)$.

The performance of a RC is directly proportional to the number of neurons in its reservoir layer [1]. There are several attempts to design hardware implementation of RC which involves multiple photonic neurons [2]. However, such an approach is not feasible for designing RCs with thousands of neurons as that would lead to large power and area overhead [3]. A solution to this problem is a single node photonic RC with delay dynamics, which is explained in the following section.

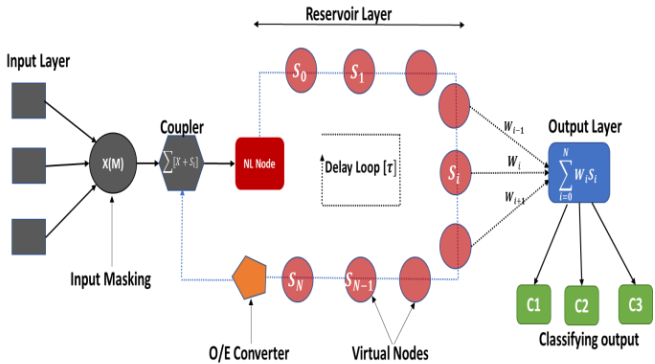

Fig.2: Single Node Photonic RC

### B. Single Node Photonic RC

A photonic implementation of RC fully exploits the advantages of optical properties of photonic hardware (low-power, high-bandwidth, and inherent parallelism). Fig.2 shows a conceptual diagram of a single node photonic RC. It consists of a non-linear node with a delay loop. An optical component which has non-linear characteristics can be used as a non-linear node e.g. microring modulator, Mach Zehnder interferometer, and optical amplifier. A spool of optical fiber can be made to provide delay. The O/E converter converts analog optical data of duration $T$ to $N$ discrete electrical samples $s_i(n)$ which is fed to the coupler one by one. Each $s_i(n)$ represents the state of a neuron in the reservoir layer at time step *n*. The non-linear node receives sum of $s_i(n)$ and masked input $x(n + 1)$ and then transforms it to $s_i(n + 1)$ as depicted in Equation.1. After that, the state value $s_i(n + 1)$ is stored in the delay loop to be used in the next time step i.e. $n + 2$. The ratio of delay $\tau$ to the O/E conversion time $t$ determines the number of neurons *N* in this kind of design.

### C. Time-shared Single Node Photonic RC

The proposed time-shared photonic RC consists of a TDM integrated input layer, a reservoir layer, and a a readout layer. A logical microarchitecture of the proposed design is depicted in Fig.3. The details are as follows.

TDM-integrated input layer: A TDM scheme is incorporated in the input layer to enable time-shared behavior and facilitate multitasking computation. For simplicity, we consider 4 tasks (p(t), q(t), r(t), s(t)) at the input to demonstrate four level multitasking although higher level multitasking is feasible. The TDM scheme involves a sample and hold circuit which samples and holds each task for an equal duration of $T_S$ in a round robin fashion. This in turn converts each continuous-time task p(t) to a discretized piecewise constant function (p(n) where $p(t) = p(n)$, $nT_S \leq t < (n + 1)T_S$), *n* is a time step. Each discrete input $p(n)$ is multiplied with a periodic mask input $m(n)$ of period $T_S$. Here $m(n) = m_i(n)$ for $(i - 1)\left(\frac{T_S}{N}\right) < n \leq (i + 1)\left(\frac{T_S}{N}\right)$; $i = 1,2,...N$; $m_i(n)$ is randomly chosen from $[-1, +1]$. The result of this multiplication is a masked input $p_i(n)$ which drives the reservoir layer. Similarly, masked input $q_i(n), r_i(n), s_i(n)$ are produced.

Reservoir layer: The reservoir layer consists of a Mach-Zehnder interferometer (MZI) as non-linear node and an optical fiber spool to provide delayed feedback.

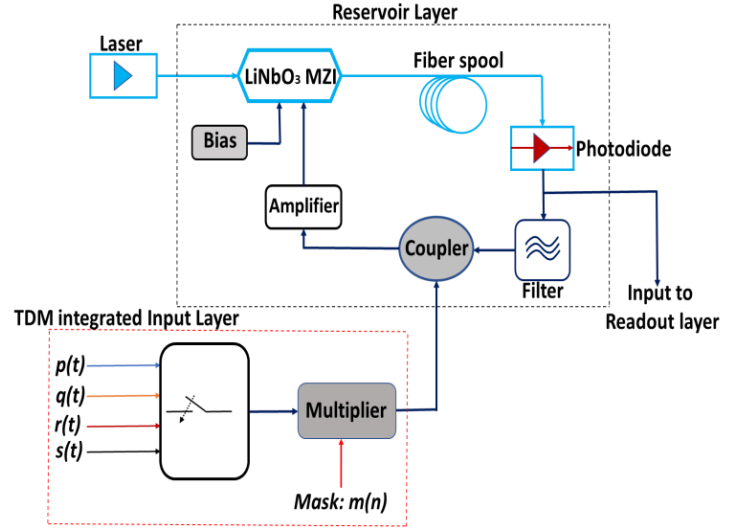

Fig.3: Time-shared Photonic RC

For each time step $n$, the photodiode in the reservoir layer converts each reservoir state $p_i(n)$ (also $q_i(n), r(n), s(n)$) from optical form to electronic form. The O/E conversion rate of the photodiode is $\frac{1}{h}$ samples per second. We consider $N$ number of reservoir states for each task. A time gap of $2h$ is introduced between consecutive tasks by the coupler to avoid signal interference. Analytically, we can write $N = \frac{T_S}{h}$. The maximum attainable value of $N$ for the proposed design is $\frac{\tau-2(k-1)h}{k}$ where $\tau$ is time delay of the fiber spool and $k$ is number of input tasks. The state of reservoir $i$ for task $p$ is given by

$$s_p^i(n) = \sin(\propto s_p^i(n-1) + \beta m^i(n)p(n) + \emptyset) \quad (3)$$

Here $\propto$ and $\beta$ are feedback gains; $\emptyset$ is a bias value; and $m^i(n)$ represents the mask input. $\propto$, $\beta$, and $\emptyset$ are adjustable parameters. The MZI used in our design has sinusoidal non-linearity; hence the above equation is based on a *sin* function.

## III. EXPERIMENTAL METHODOLOGY

We synthesize optoelectronic components such as optical fiber spool, photodiode, coupler, MZI, sampler using a commercial photonic design tool called IPKISS [4]. We use Tanner L-edit [4] to synthesize the TDM circuit. The synthesized components are used to design the proposed time-shared photonic RC system on Caphe [4]. We integrate Caphe with Oger toolbox [5] which is an open-source framework to rapidly build, train and evaluate RC models. The proposed RC system is simulated in Oger with four widely used RC benchmarks, namely spoken digit recognition, Santa Fe time series prediction, non-linear channel equalization, NARMA task.

*Spoken digit recognition*: Spoken digit recognition task is a widely-accepted big data classification task. The objective of the task is to classify ten spoken digits (0-9), each recorded ten times by five different persons. The dataset is obtained from a subset of NIST TI-46 corpus (National Institute of Standards and Technology Texas Instrument-46 corpus) [6]. The input $p_i(n)$ to the reservoir comprises of an 86 dimensional (i = 1,2,... 86) state vector with up to 130 time steps. The number of variables $N = 200$. This requires an input mask $m_{ij}$ of matrix size $N \times 86$, where each element is chosen randomly from {-1, +1} with equal probabilities. $\sum_{i=1}^{86} m_{ji}p_i(n)$ (product of input and mask) is used to drive the reservoir.

*Santa Fe time series*: A time-series is a sequence of periodic data points over a continuous time interval. In our experiment, we use Santa Fe financial time-series recorded from a far-infrared laser operating in chaotic state []. The goal of this experiment is to predict a data point one time-step ahead in the future. The dataset contains 10000 points and we use 4000 points. Prediction performance is evaluated based on the normalized mean square error (NMSE) defined as:

$$NMSE = \frac{1}{n}\sum_{i=1}^{n} \frac{(O_i' - O_i)}{\sigma O_i^2} \quad (5)$$

where $O_i'$ and $O_i$ are predicted and expected values at time step $i$, $n$ is total number of time step, and $\sigma$ is the standard deviation. Here, *NMSE = 0* implies perfect prediction and *NMSE = 1* indicates no prediction.

*Non-linear channel equalization*: Equalization of communication channel is a way to facilitate reliable wireless communication. Whenever a signal is transmitted across a wireless communication channel, it encounters noise, channel effects (e.g. distortion, dispersion), and inter-symbol interference. Equalization of a wireless communication channel has been widely used as a benchmark task for RC simulation The following two equations represent the relationship of the output s(n) a non-linear wireless channel to its input g(n).

$$z(n) = 0.08g(n+2) - 0.12g(n+1) + g(n) + 0.18g(n-1) - 0.1g(n-2) + 0.091g(n-3) - 0.05g(n-4) + 0.04gn(n-5) + 0.03g(n-6) + 0.01g(n-7) \quad (6)$$

$$s(n) = z(n) + 0.36z(n)^2 - 0.011z(n)^3 + d(n) \quad (7)$$

As depicted in Equation 7, s(n) encounters second-order $z(n)^2$ and the third-order $z(n)^3$ nonlinear distortions, and also additive Gaussian white noise d(n), which may result from the channel. s(n) is used as the final input to reservoir system.

*NARMA task*: The NARMA task is one of the most widely used benchmarks in RC. The input u(k) for this task consists of scalar random numbers, drawn from a uniform distribution in the interval [0, 0.5] and the target y( k + 1 ) is given by the following recursive formula:

$$y_{k+1} = 0.3y_k + 0.05y_k[\sum_{i=0}^{9} y_{k-i}] + 1.5u_k u_{k-9} + 0.1 \quad (8)$$

## IV. RESULTS & ANALYSIS

Using Oger, we perform timeshared computation of multiple tasks on the proposed RC. For multitasking, we consider simultaneous computations of two tasks, three tasks, and four tasks. For each case, we obtain prediction error and execution time pertaining to a task. A detailed comparison of these results with state-of-the-art RCs is as follows.

### A. Simultaneous Computation of Two Different Tasks

Here, we show the results for simultaneous prediction of two different tasks. We consider four cases, (i) Santa Fe and non-linear channel equalization, (ii) spoken digit and *NARMA*, (iii) Santa Fe and spoken digit, and (iv) non-linear channel and *NARMA*. In the first case, Santa Fe series is used as task 1 and non-linear channel equalization datasets is used as task 2. For the task 1, first 5000 data points of Santa Fe series are used for training whereas 3000 symbols of *g(n)* of non-linear channel benchmark are used for training task 2. For task 2, we choose SNR = 32 dB. Proportional to the length of datasets, task 1 is assigned $T_s = \frac{5}{8}T_{sample}$ and task 2 is assigned $T_s = \frac{3}{8}T_{sample}$. Here $T_{sample}$ represents the length of a single time step which constitutes a sample of each task for that time step. For testing, last 5000 data points from Santa Fe series and 1000 symbols of *g(n)* are used. Proportionately, task 1 is assigned $T_s = \frac{5}{6}T_{sample}$ and task 2 is assigned $T_s = \frac{1}{6}T_{sample}$. For task 1, the proposed RC predicts with an average $NMSE = 0.035 \pm 0.001$ whereas task 2 is executed with $SER = 1.38 \times 10^{-4.5}$. These results are as good as the results for single task executions in state-of-the-art RCs [1,2,3]. We perform experiments for case (ii), (iii), and (iv) with similar procedures. As shown in Fig.4, the proposed time-shared design exhibits similar performance for all cases of two-task executions (reported as P-2) compared to single task executions' results in [1,2,3].

*B. Simultaneous Computation of Three Different Tasks*

In this experiment, we carry out simultaneous prediction of three different tasks. We consider two cases here: (i) Santa Fe, non-linear channel equalization, and spoken-digit recognition, (ii) Santa-Fe, spoken-digit recognition, and NARMA. In (i), first 4000 data points of Santa Fe are used for training the Santa Fe time series, 3000 symbols of *g(n)* are used for training the non-linear channel equalization task, and 400 digits of spoken-digit datasets are used for training. Proportionately, Santa Fe task is assigned $T_s = \frac{4}{8}T_{sample}$ non-linear task is assigned $T_s = \frac{2}{8}T_{sample}$, and spoken-digit task is assigned $T_s = \frac{1}{8}T_{sample}$. During the testing phase, the last 4000 data points of Santa Fe are used for testing the Santa Fe task, 1000 symbols of *g(n)* are used for testing the non-linear channel equalization task, and 100 digits of spoken-digit datasets are used to test spoken-digit task. Our photonic RC predicts Santa Fe time series with $NMSE = 0.035 \pm 0.02$. The non-linear channel equalization task is computed with $SER = 1.25 \times 10^{-4.5}$ and the spoken-digit classification is performed with $WER = 0.08\%$. We follow similar approach to perform (ii). It is clear from the graphs in Fig.4 that the prediction error for this experiment (reported as P-3) is comparable to single task execution results in recently reported designs. However, P-3 results are slightly degraded from P-2 which is obvious.

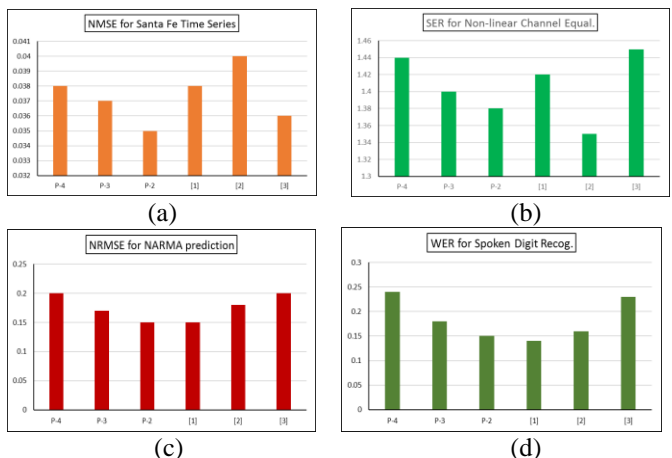

Fig.4: Prediction performance for (a) Santa Fe time series, (b) Non-linear channel equalization, (c) NARMA task, and (d) Spoken digit recognition; P-2: proposed RC with two tasks, P-3: proposed RC with three tasks, and P-4: proposed RC with four tasks.

*C. Simultaneous Computation of Four Different Tasks*

In this case, we perform time-shared computation of the four benchmarks. For training, first 3000 data points of Santa Fe series, 3000 symbols of non-linear channel, 400 digits of spoken digit task, and first 1600 points of NARMA task are sampled and fed to the photonic RC. Accordingly, these tasks are assigned sample time equals to $\frac{4}{8}th, \frac{2}{8}th, \frac{1}{8}th,$ and $\frac{1}{8}th$ of $T_{sample}$ respectively. For testing, last 1000 data points of Santa Fe series, last 1000 symbols of non-linear channel, last 400 digits of spoken digit, and last 1600 data points of NAMA task are used. Now, sample times of tasks are $\frac{2}{8}th, \frac{2}{8}th, \frac{2}{8}th,$ and $\frac{2}{8}th$ of $T_{sample}$ respectively. As shown in Fig.4, the proposed RC achieves (reported as P-4) $NMSE = 0.042 \pm 0.001$ for Santa Fe task, $SER = 1.44 \times 10^{-4.5}$ for non-linear channel equalization, $WER = 0.24\%$ for spoken digit recognition, and $NRMSE = 0.2$ for NARMA task. These results are acceptable in RC predictions though slightly degraded from the single task execution results of [1,2,3].

*D. Timing and Power Analysis*

Prediction rate of the proposed system is 3 Gigabytes per second which is same as [1-2]. Due to the integrated time-shared scheme, the proposed design computes multiple tasks in less than the time required for state-of-the-art RCs [1-4] to execute a single task. The total power consumption comprises of the power consumed by individual optoelectronic components such as TDM circuit (10 mw), MZM (5-7 W), laser and photodiode (30 W), and other components (10 W). A conservative calculation gives a total power of 50 W. Power consumption details are not provided in [1,2,3,4]. Therefore, we cannot provide a power consumption comparison.

## V. CONCLUSIONS

In this letter, we propose a single node photonic reservoir computer using delay dynamic system. We integrate a TDM circuit with RC to facilitate time-sharing of multiple tasks. We model and synthesize the proposed system using IPKISS. We integrate Oger with IPKISS framework to perform training and testing of benchmarks on the proposed RC. We execute four well known benchmarks in the proposed system and demonstrated time-shared prediction of multiple tasks with state-of-the-art performances. This design can be expanded for more than four tasks by increasing the number of input channels of the TDM circuit. Our future work in to determine the maximum number of tasks that can be computed in a time-shared fashion with reasonable performance.